\documentclass{ws}

\usepackage{amsfonts}
\usepackage{epsfig,multicol}

\newbox\mybox

\newcommand\fverb{\setbox\mybox=\hbox\bgroup\verb}
\newcommand\fverbdo{\egroup\medskip\noindent\fbox{\unhbox\mybox}\ }
\newcommand\fverbit{\egroup\item[\fbox{\unhbox\mybox}]}

\begin{document}

\title{Applications of quantum integrable systems}
\author{Olalla~A.~Castro-Alvaredo and Andreas~Fring}

\address{Institut f\"ur
Theoretische Physik, Freie Universit\"at Berlin, \\
Arnimallee 14, D-14195 Berlin, Germany \\
E-mail: olalla/fring@physik.fu-berlin.de}

\maketitle

\abstracts{We present two applications of quantum integrable systems. 
First, we predict that it is possible to generate high harmonics from
solid state devices by demonstrating  that the emission spectrum  
of a minimally coupled laser field of frequency $\omega$ to an impurity system 
of a quantum wire, contains multiples of the incoming frequency. Second, by 
evaluating expressions for the conductance in the high temperature
regime we show that the characteristic filling fractions of the Jain sequence,
which occur in the fractional quantum Hall effect, can be obtained 
from quantum wires which are described by minimal
affine Toda field theories.}

\begin{center}
\medskip
\em{Dedicated to A.A.~Belavin on his $60^{\rm th}$ birthday} 
\medskip
\end{center}

\section{Introduction}

In the context of conformal and massive integrable quantum field theories an
impressive amount of non-perturbative techniques has been developed during
the last 25 years. Needless to say that the contributions of the Landau school
has always been vital for that progress. 
The original motivation for considering such theories was
to use the lower dimensional set up as a testing ground for general
conceptual ideas and possibly to apply them in the context of string theory.
As a consequence, most of the work in this area is often of a rather formal
mathematical nature and lacks a link to direct physical application. So far
this has not been a major issue, but lately
the experimental techniques have advanced to such an extent that one can
realistically hope to measure various physical quantities which can be
predicted based on the developed approaches.

Here we want to present two examples of such quantities. 
The first is concerned with
the prediction of harmonic spectra when a three dimensional laser field is
coupled to a one dimensional quantum wire\cite{OFC}. To observe interesting
phenomena in this context one needs to consider impurity systems and the
concepts of integrability are so constraining that they lead one to consider
mostly free systems with impurities.

The second application we shall present is related to the computation of
particular values of the conductance of a quantum wire\cite{CFHall}. There
exist two established theoretical descriptions to compute the conductance,
the Kubo formula\cite{kubo,KTH}, which is the outcome of a dynamical
linear-response theory and the Landauer-B\"{u}ttinger theory\cite{Land},
which is a semi-classical transport theory. In both descriptions one can
make full use of different non-perturbative techniques developed in the 1+1
dimensional quantum field theory context. In the first approach one of the
key quantities involved is the current-current correlation function which
can be obtained from a form factor expansion\cite{KW,Smir,BCFK} and in the
second approach one requires the density distributions which are accessible
from a thermodynamic Bethe ansatz\cite{TBAZam} (TBA) analysis. Here we will
concentrate on the latter and show that the characteristic Jain filling
fractions\cite{Jain}, which occur in the fractional quantum Hall 
effect\cite{St}, 
can be obtained from quantum wires which are described by minimal
affine Toda field theories\cite{ATFT}.

\section{Harmonic generation}

We commence by briefly explaining what harmonics are. The first experimental
evidence can be traced back to the early sixties\cite{Franken}. Franken et
al found that when hitting a crystalline quartz with a weak ultraviolet
laser beam of frequency $\omega $, it emits a frequency which is $2\omega $.
Generalizing this phenomenon to higher multiples, one says nowadays that
high harmonic generation is the non-linear response of a medium (a crystal,
an atom, a gas, ...) to a laser field. The importance of harmonic generation
is related to the fact that it allows to convert infrared input radiation of
frequency $\omega $ into light in the extreme ultraviolet regime whose
frequencies are multiples of $\omega $ (even up to order $\sim 1000$, see
e.g.\cite{record} for a recent review). A typical experimental spectrum is
presented in figure 1.

\begin{figure}[h]
\epsfig{file=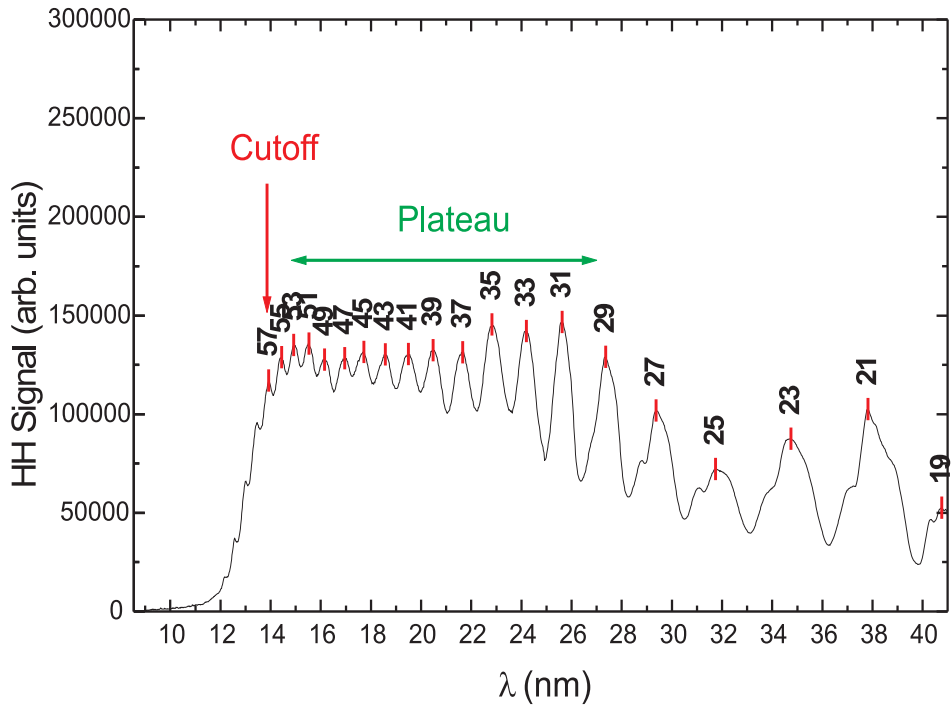,width=12.5cm,height=9.28cm} 
        \caption{Harmonic spectrum for Neon for a Ti:Sa laser with 
   $\lambda= 795nm$. 
Measured at the Max Born Institut Berlin [15].}
\end{figure}

\noindent In gases, composed of atoms or small molecules, this phenomenon is
well-understood and, to some extent, even controllable in the sense that the
frequency of the highest harmonic, the so-called \textquotedblleft
cut-off\textquotedblright , visible in figure 1, can be tuned as well as the
intensities of particular groups of harmonics. In more complex systems,
however, for instance solids, or larger molecules, high-harmonic generation
is still an open problem. This is due to the fact that, until a few years
ago, such systems were expected not to survive the strong laser fields one
needs to produce such effects. However, nowadays, with the advent of
ultrashort pulses, there exist solid-state materials whose damage threshold
is beyond the required\cite{solid1} intensities of $10^{14}\mathrm{W/cm}^{2}$.
As a direct consequence, there is an increasing interest in such
materials as potential sources for high-harmonics. In fact, several groups
are currently investigating this phenomenon in systems such as thin 
crystals\cite{mois98,thincryst2}, carbon nanotubes\cite{mois2000}, or organic
molecules\cite{benz1,mois2001}.

We will therefore try to answer here the question, whether it is possible to
generate harmonics from solid state devices and as a prototype of such a
system we study a quantum wire coupled to a laser field.

\subsection{Constraints from integrability}

In order to couple the laser field to the wire we need some seeds to attach
the field and therefore we are naturally led to consider impurity systems.
Essential quantities to compute are the transmission amplitudes. We commence
by demonstrating how integrability puts severe constraints onto them. One
exploits here one of the great advantages of integrability in 1+1
dimensional models, which is the well-known fact that the n-particle
scattering matrix factorises into two-particle S-matrices, which can be
determined by some constraining equations which are central to the entire
subject, the Yang-Baxter\cite{YB} and bootstrap equations\cite{boot}.
Similar equations hold in the presence of a boundary\cite{Chered,Skly,FK}
or a defect\cite{DMS,CFG}. It is clear that with regard to the conductance
a situation with a pure boundary, i.e.~non-trivial effects on the
constrictions, or purely transmitting defects will be rather uninteresting
and we would like to consider the case when $R$ and $T$ are simultaneously
non-vanishing. It will turn out that for that situation the Yang-Baxter
equations are so constraining that not many integrable theories will be left
to consider. Thus this section serves essentially to motivate the study of
the free Fermion, which after all is very close to a realistic system of
electrons propagating in quantum wires.

We label now particle types by Latin and degrees of freedom of the impurity
by Greek letters, the bulk scattering matrix by $S$, and the left/right
reflection and transmission amplitudes of the defect by $R/\tilde{R}$ and $T/%
\tilde{T}$, respectively. Then the transmission and reflection amplitudes
are constrained by the ``unitarity'' relations 
\begin{eqnarray}
R_{i\alpha }^{j\beta }(\theta )R_{j\beta }^{k\gamma }(-\theta )+T_{i\alpha
}^{j\beta }(\theta )\tilde{T}_{j\beta }^{k\gamma }(-\theta ) &=&\delta
_{i}^{k}\delta _{\alpha }^{\gamma },  \label{U2} \\
R_{i\alpha }^{j\beta }(\theta )T_{j\beta }^{k\gamma }(-\theta )+T_{i\alpha
}^{j\beta }(\theta )\tilde{R}_{j\beta }^{k\gamma }(-\theta ) &=&0\,,
\label{U3}
\end{eqnarray}
and the crossing-hermiticity relations 
\begin{eqnarray}
R_{_{\bar{\jmath}}}^{\alpha }(\theta ) &=&\tilde{R}_{_{\bar{\jmath}%
}}^{\alpha }(-\theta )^{\ast }=S_{j\bar{\jmath}}(2\theta )\tilde{R}%
_{j}^{\alpha }(i\pi -\theta )\,,  \label{c1} \\
T_{_{\bar{\jmath}}}^{\alpha }(\theta ) &=&\tilde{T}_{_{\bar{\jmath}%
}}^{\alpha }(-\theta )^{\ast }=\tilde{T}_{j}^{\alpha }(i\pi -\theta )\,.
\label{c2}
\end{eqnarray}
The equations (\ref{U2}) and (\ref{U3}) also hold after performing a parity
transformation, that is for $R\leftrightarrow \tilde{R}$ and $%
T\leftrightarrow \tilde{T}$.

Depending now on the choice of the initial asymptotic condition one can
derive the following two non-equivalent sets of generalized Yang-Baxter
equations by exploiting the associativity of the extended
Zamolodchikov-Faddeev algebra\cite{Chered,Skly,FK,DMS,CFG} 
\begin{eqnarray}
S(\theta _{12})[\mathbb{I}\otimes R_{\alpha }^{\beta }(\theta _{1})]S(\hat{%
\theta}_{12})[\mathbb{I}\otimes R_{\beta }^{\gamma }(\theta _{2})] &=&[%
\mathbb{I}\otimes R_{\alpha }^{\beta }(\theta _{2})]S(\hat{\theta}_{12})[%
\mathbb{I}\otimes R_{\beta }^{\gamma }(\theta _{1})]S(\theta _{12}),
\label{YBt1} \\
S(\theta _{12})[\mathbb{I}\otimes R_{\alpha }^{\beta }(\theta _{1})]S(\hat{%
\theta}_{12})[\mathbb{I}\otimes T_{\beta }^{\gamma }(\theta _{2})]
&=&R_{\beta }^{\gamma }(\theta _{1})\otimes T_{\alpha }^{\beta }(\theta
_{2}),  \label{YBt2} \\
S(\theta _{12})[T_{\alpha }^{\beta }(\theta _{2})\otimes T_{\beta }^{\gamma
}(\theta _{1})] &=&[T_{\alpha }^{\beta }(\theta _{1})\otimes T_{\beta
}^{\gamma }(\theta _{2})]S(\theta _{12}),  \label{YBt3}
\end{eqnarray}
and 
\begin{eqnarray}
R_{\alpha }^{\beta }(\theta _{1})\otimes \tilde{R}_{\beta }^{\gamma }(\theta
_{2}) &=&R_{\beta }^{\gamma }(\theta _{1})\otimes \tilde{R}_{\alpha }^{\beta
}(\theta _{2}),  \label{RR} \\
\lbrack T_{\alpha }^{\beta }(\theta _{2})\otimes \mathbb{I}]S(\hat{\theta}%
_{12})[\tilde{R}_{\beta }^{\gamma }(\theta _{1})\otimes \mathbb{I}]S(\theta
_{12}) &=&T_{\beta }^{\gamma }(\theta _{2})\otimes \tilde{R}_{\alpha
}^{\beta }(\theta _{1}),  \label{TR} \\
\lbrack \mathbb{I}\otimes \tilde{T}_{\alpha }^{\beta }(\theta _{2})]S(\hat{%
\theta}_{12})[\mathbb{I}\otimes R_{\beta }^{\gamma }(\theta _{1})]S(\theta
_{12}) &=&R_{\alpha }^{\beta }(\theta _{1})\otimes \tilde{T}_{\beta
}^{\gamma }(\theta _{2}),  \label{RT} \\
\lbrack T_{\alpha }^{\beta }(\theta _{1})\otimes \mathbb{I}]S(\hat{\theta}%
_{12})[\tilde{T}_{\beta }^{\gamma }(\theta _{2})\otimes \mathbb{I}] &=&[%
\mathbb{I}\otimes \tilde{T}_{\alpha }^{\beta }(\theta _{2})]S(\hat{\theta}%
_{12})[\mathbb{I}\otimes T_{\beta }^{\gamma }(\theta _{1})].  \label{TT}
\end{eqnarray}
We used here the convention $(A\otimes B)_{ij}^{kl}=A_{i}^{k}B_{j}^{l}$ for
the tensor product and abbreviated the rapidity sum $\hat{\theta}%
_{12}=\theta _{1}+\theta _{2}$ and difference $\theta _{12}=\theta
_{1}-\theta _{2}$. Once again the same equations also hold for $%
R\leftrightarrow \tilde{R}$ and $T\leftrightarrow \tilde{T}$.

Apart from some discrepancies in the indices the equations (\ref{YBt1})-(\ref%
{YBt3}) correspond to a more simplified, in the sense that there were no
degrees of freedom in the defect and parity invariance is assumed, set of
equations considered previously in\cite{DMS}. For diagonal scattering it
was argued in\cite{DMS} that one can only have reflection and transmission
simultaneously when $S=\pm 1$. In\cite{CFG} a more general set up which
includes all degrees of freedom was studied. A second set of equations (\ref%
{RR})-(\ref{TT}), which is not equivalent to (\ref{YBt1})-(\ref{YBt3}) was
found. It was shown that in the absence of degrees of freedom in the defect
no theory which has a non-diagonal bulk scattering matrix admits
simultaneous reflection and transmission. This result even holds for the
completely general case including degrees of freedom in the defect upon a
mild assumption on the commutativity of $R$ and $T$ in these variables. It
was further shown that besides $S=\pm 1$ also the Federbush model\cite%
{Feder} and the generalized coupled Federbush models\cite{Fform} allow for $%
R\neq 0$ and $T\neq 0$. However, when treating non-relativistic theories,
the amplitutes depend only on the individual rapidities, 
such that one can find non-trivial solutions 
with $R\neq 0$ and $T\neq 0$ even when the bulk theory is not free\cite{MRS}.

\subsection*{2.1.1 Multiple impurity systems}

The most interest situation in impurity systems arises when instead of a
single one considers multiple defects, since that leads to the occurrence of
resonance phenomena and when the number of defects tends to infinity even to
band structures. Assuming that the distance between the defects is small in
comparison to the length of the wire one can easily construct the
transmission and reflection amplitudes of the multiple defect system from
the knowledge of the corresponding quantities in the single defect system.
For instance for two defects one obtains 
\begin{eqnarray}
T_{i}^{\alpha \beta }(\theta ) &=&\frac{T_{i}^{\alpha }(\theta )T_{i}^{\beta
}(\theta )}{1-R_{i}^{\beta }(\theta )\tilde{R}_{i}^{\alpha }(\theta )}%
,\qquad R_{i}^{\alpha \beta }(\theta )=R_{i}^{\alpha }(\theta )+\frac{%
R_{i}^{\beta }(\theta )T_{i}^{\alpha }(\theta )\tilde{T}_{i}^{\alpha
}(\theta )}{1-R_{i}^{\beta }(\theta )\tilde{R}_{i}^{\alpha }(\theta )},
\label{tr} \\
\tilde{T}_{i}^{\alpha \beta }(\theta ) &=&\frac{\tilde{T}_{i}^{\alpha
}(\theta )\tilde{T}_{i}^{\beta }(\theta )}{1-R_{i}^{\beta }(\theta )\tilde{R}%
_{i}^{\alpha }(\theta )},\qquad \tilde{R}_{i}^{\alpha \beta }(\theta )=%
\tilde{R}_{i}^{\beta }(\theta )+\frac{R_{i}^{\alpha }(\theta )T_{i}^{\beta
}(\theta )\tilde{T}_{i}^{\beta }(\theta )}{1-R_{i}^{\beta }(\theta )\tilde{R}%
_{i}^{\alpha }(\theta )}.  \label{tr2}
\end{eqnarray}
These expressions allow for a direct intuitive understanding, for instance
we note that the term $[1-R_{i}^{\beta }(\theta )\tilde{R}_{i}^{\alpha
}(\theta )]^{-1}=\sum_{n=1}^{\infty }(R_{i}^{\beta }(\theta )\tilde{R}%
_{i}^{\alpha }(\theta ))^{n}$ simply results from the infinite number of
reflections which we have in-between the two defects. This is of course well
known from Fabry-Perot type devices of classical and quantum optics. For the
case $T=\tilde{T},R=\tilde{R}$ the expressions (\ref{tr}) and (\ref{tr2})
coincide with the formulae proposed in\cite{Konik2}. When absorbing the
space dependent phase factor into the defect matrices, the explicit example
presented in\cite{DMS} for the free Fermion perturbed with the energy
operator agree almost for $T=\tilde{T},R=\tilde{R}$ \ with the general
formulae (\ref{tr}). They disagree in the sense that the equality of $%
R_{i}^{\alpha \beta }(\theta )$ and $\tilde{R}_{i}^{\alpha \beta }(\theta )$
does not hold for generic $\alpha ,\beta $ as stated in\cite{DMS}.

It is now straightforward to generalize the expressions for an arbitrary
number of defects, say $n$, in a recursive manner 
\begin{eqnarray}
T_{i}^{\vec{\alpha}}(\theta ) &=&\frac{T_{i}^{\alpha _{1}\ldots \alpha
_{k}}(\theta )T_{i}^{\alpha _{k+1}\ldots \alpha _{n}}(\theta )}{1-\tilde{R}%
_{i}^{\alpha _{1}\ldots \alpha _{k}}(\theta )R_{i}^{\alpha _{k+1}\ldots
\alpha _{n}}(\theta )},\qquad \,\,\qquad \qquad \qquad
\,\,\,\,\,\,\,\,\,\,\,1<k<n\,,  \label{ttr} \\
R_{i}^{\vec{\alpha}}(\theta ) &=&R_{i}^{\alpha _{1}\ldots \alpha
_{k}}(\theta )+\frac{R_{i}^{\alpha _{k+1}\ldots \alpha _{n}}(\theta
)T_{i}^{\alpha _{1}\ldots \alpha _{k}}(\theta )\tilde{T}_{i}^{\alpha
_{1}\ldots \alpha _{k}}(\theta )}{1-\tilde{R}_{i}^{\alpha _{1}\ldots \alpha
_{k}}(\theta )R_{i}^{\alpha _{k+1}\ldots \alpha _{n}}(\theta )},\quad
\,\,1<k<n\,.\,\,  \label{ttr2}
\end{eqnarray}
We encoded here the defect degrees of freedom into the vector $\vec{\alpha}%
\mathbf{=}\{\alpha _{1},\cdots ,\alpha _{n}\}$. Similar expressions also
hold for $\tilde{T}_{i}^{\vec{\alpha}}(\theta )=\tilde{T}_{i}^{\alpha
_{1}\ldots \alpha _{n}}(\theta )$ and $\tilde{R}_{i}^{\vec{\alpha}}(\theta )=%
\tilde{R}_{i}^{\alpha _{1}\ldots \alpha _{n}}(\theta )$.

Alternatively, we can define, in analogy to standard quantum mechanical
methods (see e.g.\cite{CT}), a transmission matrix which takes the particle 
$i$ from one side of the defect of type $\alpha $ to the other 
\begin{equation}
\mathcal{M}_{\alpha }^{i}(\theta )=\left( 
\begin{array}{cc}
T_{i}^{\alpha }(\theta )^{-1} & -R_{i}^{\alpha }(\theta )T_{i}^{\alpha
}(\theta )^{-1} \\ 
-R_{i}^{\alpha }(-\theta )T_{i}^{\alpha }(-\theta )^{-1} & T_{i}^{\alpha
}(-\theta )^{-1}%
\end{array}
\right) \,.
\end{equation}
Then alternatively to the recursive way (\ref{ttr}) and (\ref{ttr2}), we can
also compute the multi-defect transmission and reflection amplitudes as 
\begin{equation}
T_{i}^{\vec{\alpha}}(\theta )=\left( \prod_{k=1}^{n}\mathcal{M}_{\alpha
_{k}}^{i}(\theta )\right) _{11}^{-1},\,\,\,\quad R_{i}^{\vec{\alpha}}(\theta
)=-\left( \prod_{k=1}^{n}\mathcal{M}_{\alpha _{k}}^{i}(\theta )\right)
_{12}\left( \prod_{k=1}^{n}\mathcal{M}_{\alpha _{k}}^{i}(\theta )\right)
_{11}^{-1}.  \label{ttr3}
\end{equation}
This formulation has the virtue that it is more suitable for numerical
computations, since it just involves matrix multiplications rather than
recurrence operations. In addition, it allows for an elegant analytical
computation of the band structures for $n\rightarrow \infty $, which we will
however not comment upon any further.

\subsection{Constraints from potential scattering theory}

\noindent As we argued above, in order to obtain a non-trivial conductance
we are lead by integrability to consider free theories, 
possibly with some exotic statistics.
Trying to be as close as possible to some realistic situation, 
i.e.~electrons, we consider first the free Fermion, 
which, with a line of defect,
was first treated in\cite{Cabra}. Thereafter it has also been considered 
in\cite{GZ,DMS} and\cite{Konik} from different points of view. In\cite%
{Cabra,GZ,DMS} the defect line was taken to be of the form of the energy
operator and in\cite{Konik} also a perturbation in form of a single Fermion
has been considered. In\cite{OA45} we treated a much wider class of
possible defects.

Let us consider the Lagrangian density for a complex free Fermion $\psi $
with $\ell $ defects\footnote{%
We use the conventions: 
\begin{eqnarray*}
x^{\mu } &=&(x^{0},x^{1}),\qquad p^{\mu }=(m\cosh \theta ,m\sinh \theta
),\quad g^{00}=-g^{11}=\varepsilon ^{01}=-\varepsilon ^{10}=1, \\
\gamma ^{0} &=&\left( 
\begin{array}{cc}
0 & 1 \\ 
1 & 0%
\end{array}
\right) ,\quad \quad \gamma ^{1}=\left( 
\begin{array}{cc}
0 & 1 \\ 
-1 & 0%
\end{array}
\right) ,\quad \gamma ^{5}=\gamma ^{0}\gamma ^{1},\quad \quad \psi _{\alpha
}=\left( 
\begin{array}{c}
\psi _{\alpha }^{(1)} \\ 
\psi _{\alpha }^{(2)}%
\end{array}
\right) ,\quad \bar{\psi}_{\alpha }=\psi _{\alpha }^{\dagger }\gamma ^{0}\,.
\end{eqnarray*}%
\par
We adopt relativistic units $1=c=\hbar =m\approx e^{2}137$ as mostly used in
the particle physics context rather than atomic units $1=e=\hbar =m\approx
c/137$ more natural in atomic physics.} 
\begin{equation}
\mathcal{L}=\bar{\psi}(i\gamma ^{\mu }\partial _{\mu }-m)\psi
\,+\sum_{n=1}^{\ell }\mathcal{D}^{\alpha _{n}}(\bar{\psi},\psi ,\partial _{t}%
\bar{\psi},\partial _{t}\psi )\delta (x-x_{n})\,.  \label{LF}
\end{equation}
The defect is described here by the functions $\mathcal{D}^{\alpha _{n}}(%
\bar{\psi},\psi ,\partial _{t}\bar{\psi},\partial _{t}\psi )$, which we
assume to be linear in the Fermi fields $\bar{\psi}$,$\psi $ and their time
derivatives. We can now proceed in analogy to standard quantum mechanical
potential scattering theory (see also\cite{GZ,DMS,Konik}) and construct the
amplitudes by adequate matching conditions on the field. We consider first a
single defect at the origin which suffices, since multiple defect amplitudes
can be constructed from the single defect ones, according to the arguments
of the previous section. We decompose the fields of the bulk theory as $\psi
(x)=\Theta (x)$ $\psi _{+}(x)+\Theta (-x)$ $\psi _{-}(x)$, with $\Theta (x)$
being the Heavyside unit step function, and substitute this ansatz into the
equations of motion. As a matching condition we read off the factors of the
delta function and hence obtain the constraints 
\begin{equation}
i\gamma ^{1}(\psi _{+}(x)-\psi _{-}(x))|_{x=0}=\left. \frac{\partial 
\mathcal{D}}{\partial \bar{\psi}(x)}\right\vert _{x=0}\,\,-\left. \frac{%
\partial }{\partial t}\left[ \frac{\partial \mathcal{D}}{\partial (\partial
_{t}\bar{\psi}(x))}\right] \right\vert _{x=0}.  \label{bbcon}
\end{equation}
We then use for the left ($-$) and right ($+$) parts of $\psi $ the
well-known Fourier decomposition of the free field 
\begin{equation}
\psi _{j}^{f}(x)=\int \frac{d\theta }{\sqrt{4\pi }}\left( a_{j}(\theta
)u_{j}(\theta )e^{-ip_{j}\cdot x}+a_{_{\bar{\jmath}}}^{\dagger }(\theta
)v_{j}(\theta )e^{ip_{j}\cdot x}\right) \,,\qquad  \label{free}
\end{equation}
with the Weyl spinors 
\begin{equation}
u_{j}(\theta )=-i\gamma ^{5}v_{j}(\theta )=\sqrt{\frac{m_{j}}{2}}\left( 
\begin{array}{c}
e^{-\theta /2} \\ 
e^{\theta /2}%
\end{array}
\right) \mathrm{\,\,}  \label{WS}
\end{equation}
and substitute them into the constraint (\ref{bbcon}). Treating the
equations obtained in this manner componentwise, stripping off the
integrals, one can bring them thereafter into the form 
\begin{equation}
a_{j,-}(\theta )=R_{_{j}}(\theta )a_{j,-}(-\theta )+T_{_{j}}(\theta
)a_{j,+}(\theta )\,\,,
\end{equation}
which defines the reflection and transmission amplitudes in an obvious
manner. When parity invariance is broken, the corresponding amplitudes from
the right to the left do not have to be identical and we also have 
\begin{equation}
a_{j,+}(-\theta )=\tilde{T}_{_{j}}(\theta )a_{j,-}(-\theta )+\tilde{R}%
_{j}(\theta )a_{j,+}(\theta )\,.
\end{equation}
The creation and annihilation operators $\,a_{i}^{\dagger }(\theta )$ and $%
a_{i}(\theta )$ satisfy the usual fermionic anti-commutation relations $%
\{a_{i}(\theta _{1}),a_{j}(\theta _{2})\}=0$, $\{a_{i}(\theta
_{1}),a_{j}^{\dagger }(\theta _{2})\}=2\pi \delta _{ij}\delta (\theta _{12})$%
. In this way one may construct the $R$'s and $T$'s for any concrete defect
which is of the generic form as described in (\ref{LF}). After the
construction one may convince oneself that the expressions found this way
indeed satisfy the consistency equations like unitarity (\ref{U2}), (\ref{U3}%
) and crossing (\ref{c1}), (\ref{c2}). Unfortunately the equations (\ref{U2}%
)-(\ref{c2}) can not be employed for the construction, since they are not
restrictive enough by themselves to determine the $R$'s and $T$'s. We
consider now some concrete example:

\subsection*{2.2.1 Impurities 
of Luttinger liquid type $\mathcal{D}(\bar{\protect%
\psi},\protect\psi )=\bar{\protect\psi}(g_{1}+g_{2}\protect\gamma ^{0})%
\protect\psi $}

Luttinger liquids\cite{Lutt} are of great
interest in condensed matter physics, which is one of the motivations for
our concrete choice of the defect $\mathcal{D}(\bar{\psi},\psi )=\bar{\psi}%
(g_{1}+g_{2}\gamma ^{0})\psi $. When taking the conformal limit of the
defect one obtains an impurity which played a role in this context, see 
e.g.\cite{Aff}, after eliminating the bosonic number counting operator. In the
way outlined above, we compute the related transmission and reflection
amplitudes 
\begin{eqnarray}
R_{j}(\theta ,g_{1},g_{2},-y) &=&\tilde{R}_{j}(\theta ,g_{1},g_{2},y)=\frac{%
4i(g_{2}+g_{1}\cosh \theta )e^{2iym\sinh \theta }}{(4+g_{1}^{2}-g_{2}^{2})%
\sinh \theta -4i(g_{1}+g_{2}\cosh \theta )}\,, \\
R_{\bar{\jmath}}(\theta ,g_{1},g_{2},-y) &=&\tilde{R}_{\bar{\jmath}}(\theta
,g_{1},g_{2},y)=\frac{4i(g_{1}-g_{2}\cosh \theta )e^{-2iym\sinh \theta }}{%
(4+g_{1}^{2}-g_{2}^{2})\sinh \theta -4i(g_{1}-g_{2}\cosh \theta )}\,, \\
T_{j}(\theta ,g_{1},g_{2}) &=&\tilde{T}_{j}(\theta ,g_{1},g_{2})=\frac{%
(4+g_{2}^{2}-g_{1}^{2})\sinh \theta }{(4+g_{1}^{2}-g_{2}^{2})\sinh \theta
-4i(g_{1}+g_{2}\cosh \theta )}\,\,, \\
T_{\bar{\jmath}}(\theta ,g_{1},g_{2}) &=&\tilde{T}_{\bar{\jmath}}(\theta
,g_{1},g_{2})=\frac{(4+g_{2}^{2}-g_{1}^{2})\sinh \theta }{%
(4+g_{1}^{2}-g_{2}^{2})\sinh \theta -4i(g_{1}-g_{2}\cosh \theta )}\,.
\end{eqnarray}
In the limit $\lim_{g_{2}\rightarrow 0}\mathcal{D}(\bar{\psi},\psi )=g_{1}%
\bar{\psi}\psi $, we recover the related results for the $T/\tilde{T}$'s and 
$R/\tilde{R}$'s for the energy defect operator. For this type of defect we
present $|T|^{2}$ and $|R|^{2}$ in figure 2 with varying parameters in order
to illustrate some of the characteristics of these functions.

Part (a) of figure 2 confirms the unitarity relation (\ref{U2}). Part (b)
and (c) show the typical resonances of a double defect, which become
stretched out and pronounced with respect to the energy when the distance
becomes smaller and the coupling constant increases, respectively. Part (d)
exhibits a general feature, that is when the number of defects is increased,
for fixed distance between the outermost defects, the resonances become more
and more dense in that region such that one may speak of energy bands.

\begin{figure}[h]
\epsfig{file=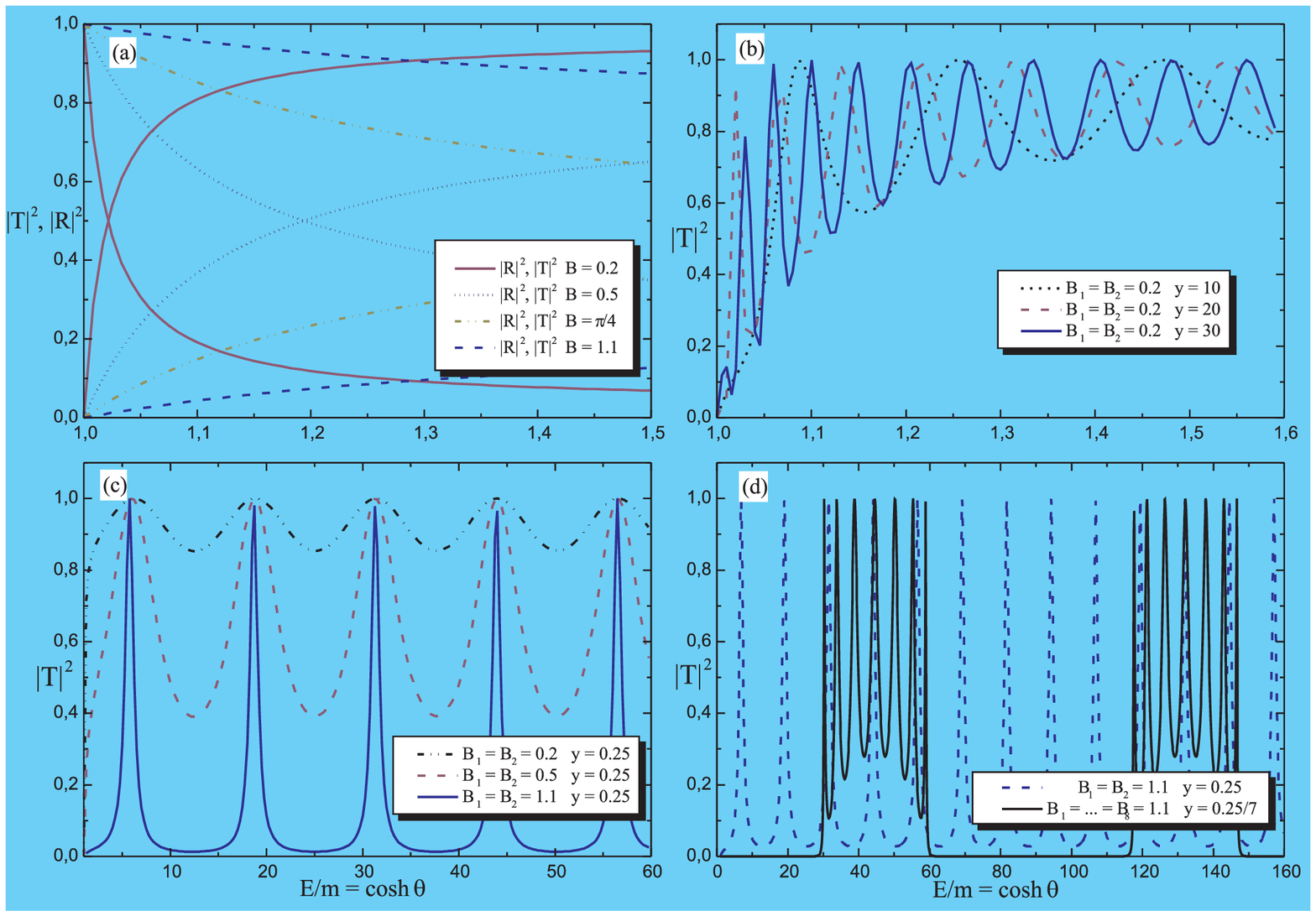,width=12.5cm,height=9.28cm} 
        \caption{  (a) Single defect with varying coupling
constant. $|T|^{2}$ and $|R|^{2}$ correspond to curves starting at 0 and 1
of the same line type, respectively. (b) Double defect with varying distance 
$y$ . (c) Double defect with varying effective coupling constant 
$B=$ arcsin$(-4g_1/(4+g_1^2))  $. 
(d) Double defect $\equiv $  dotted line, eight defects $\equiv $
solid line.} 
\end{figure}

\subsection{Laser fields interacting with quantum wires}

Let us now consider a more complex situation in which a three dimensional
laser field hits the quantum wire polarized in such a way that it has a
vector field component along the wire. Since the work of Weyl\cite{Weyl},
one knows that matter may be coupled to light by means of a local gauge
transformation, which reflects itself in the usual minimal coupling
prescription, i.e.~$\partial _{\mu }\rightarrow \partial _{\mu }-ieA_{\mu }$%
, with $A_{\mu }$ being the vector gauge potential. The free Fermions in the
wire are then described by the Lagrangian density 
\begin{equation}
\mathcal{L}_{A}=\bar{\psi}(i\gamma ^{\mu }\partial _{\mu }-m+e\gamma ^{\mu
}A_{\mu })\psi \,.  \label{La}
\end{equation}
When the laser field is switched on, we can solve the equation of motion
associated to (\ref{La}) 
\begin{equation}
(i\gamma ^{\mu }\partial _{\mu }-m+e\gamma ^{\mu }A_{\mu })\psi =0
\end{equation}
by a Gordon-Volkov type solution\cite{GV} 
\begin{equation}
\psi _{j}^{A}(x,t)=\exp \left[ ie\int^{x}dsA_{1}(s,t)\right] \psi
_{j}^{f}(x,t)=\exp \left[ ie\int^{t}dsA_{0}(x,s)\right] \psi _{j}^{f}(x,t)\,.
\label{lo2}
\end{equation}
Using now a linearly polarized laser field along the direction of the wire,
the vector potential can typically be taken in the dipole approximation to
be a superposition of monochromatic light with frequency $\omega $, i.e. 
\begin{equation}
A(t):=A_{1}(t)=\frac{1}{x}\int_{0}^{t}dsA_{0}(s)=-\frac{1}{2}%
\int_{0}^{t}dsE(s)=-\frac{E_{0}}{2}\int_{0}^{t}dsf(s)\cos (\omega s)
\label{EF1}
\end{equation}
with $f(t)$ being an arbitrary enveloping function equal to zero for $t<0$
and $t>\tau $, such that $\tau $ denotes the pulse length. In the following
we will always take $f(t)=\Theta (t)\Theta (\tau -t)\,$,$\ $with $\Theta (x)$
being again the Heavyside unit step function. The second equality in (\ref%
{EF1}), $A_{0}(x,t)=x\dot{A}(t)$, follows from the fact that we have to
solve (\ref{lo2}).

We want to comment on the validity of the dipole approximation in this
context. It consists usually in neglecting the spatial dependence of the
laser field, which is justified when $x\omega <c=1$, where $x$ is a
representative scale of the problem considered. In the context of atomic
physics this is typically the Bohr radius. In the problem investigated here,
this approximation has to hold over the full spatial range in which the
Fermion follows the electric field. We can estimate this classically, in
which case the maximal amplitude is $eE_{0}/\omega ^{2}$ and therefore the
following constraint has to hold 
\begin{equation}
\left( \frac{eE_{0}}{\omega }\right) ^{2}=4U_{p}<1\,,  \label{Dipole}
\end{equation}
for the dipole approximation to be valid. Due to the fact that $x$ is a
function of $\omega $, we have now a lower bound on the frequency rather
than an upper one as is more common in the context of atomic physics. We
have also introduced here the ponderomotive energy $U_{p}$ for monochromatic
light, that is the average kinetic energy transferred from the laser field
to the electron in the wire.

The solutions to the equations of motion of the free system and the one
which includes the laser field are then related by a factor similar to the
gauge transformation from the length to the velocity gauge 
\begin{equation}
\psi _{j}^{A}(x,t)=\exp \left[ ixeA(t)\right] \psi _{j}^{f}(x)\,.
\label{LLa}
\end{equation}
In an analogous fashion one may use the same minimal coupling procedure also
to couple in addition the laser field to the defect. One has to invoke the
equation of motion in order to carry this out. For convenience we assume now
that the defect is linear in the fields $\bar{\psi}$ and $\psi $. The
Lagrangian density for a complex free Fermion $\psi $ with $\ell $ defects $%
\mathcal{D}^{\alpha }(\bar{\psi},\psi ,A_{\mu })$ of type $\alpha $ at the
position $x_{n}$ subjected to a laser field then reads 
\begin{equation}
\mathcal{L}_{AD}=\mathcal{L}_{A}+\sum_{n=1}^{\ell }\mathcal{D}^{\alpha _{n}}(%
\bar{\psi},\psi ,A_{\mu })\,\delta (x-x_{n})\,.  \label{Lda}
\end{equation}
Considering for simplicity first the case of a single defect situated at $%
x=0 $, the solution to the equation of motion resulting from (\ref{Lda}) is
taken to be of the form $\psi _{j}^{A}(x,t)=\Theta (x)\psi
_{j,+}^{A}(x,t)+\Theta (-x)\psi _{j,-}^{A}(x,t)\,$, which means as before we
distinguish here by notation the solutions (\ref{LLa}) on the left and right
of the defect, $\psi _{j,-}^{A}(x,t)$ and $\psi _{j,+}^{A}(x,t)$,
respectively. Proceeding as before, the matching condition reads now 
\begin{equation}
i\gamma ^{1}(\psi _{j,+}^{A}(x,t)-\psi _{j,-}^{A}(x,t))|_{x=0}=\left. \frac{%
\partial \mathcal{D}_{AD}(\bar{\psi},\psi ,A_{\mu })}{\partial \bar{\psi}%
_{j}^{A}(x,t)}\right| _{x=0}.  \label{bcon}
\end{equation}

\noindent It is clear, that in this case the transmission and reflection
amplitudes will in addition to $\theta $ and $g$ also depend on the
characteristic parameters of the laser field 
\begin{equation}
T(\theta ,g,E_{0},\omega ,t)\quad \quad \mathrm{and\qquad }R(\theta
,g,E_{0},\omega ,t)\,.  \label{TRo}
\end{equation}
It is clear that the laser field can be used to control the conductance. For
instance defects which have transmission amplitudes of the form as the solid
line in figure 1 (c), can be used as optically controllable switching
devices. We can now turn to the central question which we address here,
namely whether it is possible to generate harmonics in quantum wires.

\subsection{Analysis of the transmission amplitudes}

In order to answer that question, we first have to study the spectrum of
frequencies which is filtered out by the defect while the laser pulse is
non-zero. The Fourier transforms of the reflection and transmission
probabilities provide exactly this information 
\begin{eqnarray}
\mathcal{T}(\Omega ,\theta ,E_{0},\omega ,\tau ) &=&\frac{1}{\tau }%
\int_{0}^{\tau }dt|T(\theta ,E_{0},\omega ,t)|^{2}\cos (\Omega t),
\label{TTT} \\
\mathcal{R}(\Omega ,\theta ,E_{0},\omega ,\tau ) &=&\frac{1}{\tau }%
\int_{0}^{\tau }dt|R(\theta ,E_{0},\omega ,t)|^{2}\cos (\Omega t).
\end{eqnarray}
When parity is preserved for the reflection and transmission amplitudes,
that is for real defects with $\mathcal{D}^{\ast }=\mathcal{D}$, we have $%
|T|^{2}+|R|^{2}=1$, and it suffices to consider $\mathcal{T}$ \ in the
following.

\subsection*{2.4.1 Type I defects}

Many features can be understood analytically. Taking the laser field in form
of monochromatic light in the dipole approximation (\ref{EF1}), we may
naturally assume that the transmission probability for some particular
defects can be expanded as 
\begin{equation}
|T_{I}(\theta ,U_{p},\omega ,t)|^{2}=\sum_{k=0}^{\infty }t_{2k}(\theta
)(4U_{p})^{k}\sin ^{2k}(\omega t).  \label{exp}
\end{equation}
We shall refer to defects which admit such an expansion as ``type I
defects''. Assuming that the coefficients $t_{2k}(\theta )$ become at most $%
1 $, we have to restrict our attention to the regime $4U_{p}<1$ in order for
this expansion to be meaningful for all $t$. Note that this is no further
limitation, since it is precisely the same constraint as already encountered
for the validity of the dipole approximation (\ref{Dipole}). The functional
dependence of (\ref{exp}) will turn out to hold for various explicit defects
considered below. Based on this equation, we compute for such type of defect 
\begin{equation}
\mathcal{T}_{I}(\Omega ,\theta ,U_{p},\omega ,\tau )=\sum_{k=0}^{\infty }%
\frac{(2k)!(U_{p})^{k}\sin (\tau \Omega )t_{2k}(\theta )}{\tau \Omega
\prod_{l=1}^{k}[l^{2}-(\Omega /2\omega )^{2}]}\,.  \label{ts}
\end{equation}
It is clear from this expression that type I defects will preferably let
even multiples of the basic frequency $\omega $ pass, whose amplitudes will
depend on the coefficients $t_{2k}(\theta )$. When we choose the pulse
length to be integer cycles, i.e.~$\tau =2\pi n/\omega $ for $n\in \mathbb{Z}
$, the expression in (\ref{ts}) reduces even further. The values at even
multiples of the basic frequency are simply 
\begin{equation}
\mathcal{T}_{I}(2n\omega ,\theta ,U_{p})=(-1)^{n}\sum_{k=0}^{\infty
}t_{2k}(\theta )\left( U_{p}\right) ^{k}\left( 
\begin{array}{c}
2k \\ 
k-n%
\end{array}
\right) ,  \label{tss}
\end{equation}
which becomes independent of the pulse length $\tau $. Notice also that the
dependence on $E_{0}$ and $\omega $ occurs in the combination of the
ponderomotive energy $U_{p}$. Further statements require the precise form of
the coefficients $t_{2k}(\theta )$ and can only be made with regard to a
more concrete form of the defect.

\subsection*{2.4.2 Type II defects}

Clearly, not all defects are of the form (\ref{exp}) and we have to consider
also expansions of the type 
\begin{equation}
|T_{II}(\theta ,E_{0}/e,\omega ,t)|^{2}=\sum_{k,p=0}^{\infty
}t_{2k}^{p}(\theta )\frac{E_{0}^{2k+p}}{\omega ^{2k}}\cos ^{p}(\omega t)\sin
^{2k}(\omega t).  \label{exp2}
\end{equation}
We shall refer to defects which admit such an expansion as ``type II
defects''. In this case we obtain 
\begin{eqnarray}
\mathcal{T}_{II}(\Omega ,\theta ,E_{0}/e,\omega ,\tau )
&=&\sum_{k,p=0}^{\infty }\sum_{l=0}^{p}\left( 
\begin{array}{c}
p \\ 
l%
\end{array}
\right) \frac{\Omega \sin (\tau \Omega )}{(-1)^{l+1}\tau \omega ^{2+2k}}%
E_{0}^{2k+2p}  \nonumber \\
&&\times \left( \frac{(2k+2l)!t_{2k}^{2p}(\theta )}{\prod%
\limits_{q=0}^{k+l}[(2q)^{2}-(\frac{\Omega }{\omega })^{2}]}+\frac{%
(2k+2l)!t_{2k}^{2p+1}(\theta )E_{0}}{\prod\limits_{q=1}^{k+l+1}[(2q-1)^{2}-(%
\frac{\Omega }{\omega })^{2}]}\right) .
\end{eqnarray}
We observe from this expression that type II defects will filter out all
multiples of $\omega $. For the pulse being once again of integer cycle
length, this reduces to 
\begin{equation}
\mathcal{T}_{II}(2n\omega ,\theta ,U_{p},E_{0})=\sum_{k,p=0}^{\infty
}\sum_{l=0}^{p}\frac{(-1)^{l+n} t_{2k}^{2p}(\theta )}{2^{2l-2p}}\left(
U_{p}\right) ^{k+p}E_{0}^{2p}\left( 
\begin{array}{c}
p \\ 
l%
\end{array}
\right) \left( 
\begin{array}{c}
2k+2l \\ 
k+l-n%
\end{array}
\right)  \label{II1}
\end{equation}
and 
\begin{eqnarray}
\mathcal{T}_{II}((2n-1)\omega ,\theta ,E_{0}/e) &=&\sum_{k,p=0}^{\infty
}\sum_{l=0}^{p}(-1)^{l+n+1}\frac{t_{2k}^{2p+1}(\theta )}{2^{2l-2p+1}}\left(
U_{p}\right) ^{k+p}  \nonumber \\
&&\times \left( 
\begin{array}{c}
p \\ 
l%
\end{array}
\right) \frac{(2k+2l)!(2n-1)E_{0}^{2p+1}}{(l+k-n+1)!(l+n+k)!},  \label{II2}
\end{eqnarray}
which are again independent of $\tau $. We observe that in this case we can
not combine the $E_{0}$ and $\omega $ into a $U_{p}$.

\subsection{One particle approximation}

In spite of the fact that we are dealing with a quantum field theory, it is
known that a one particle approximation to the Dirac equation is very useful
and physically sensible when the external forces vary only slowly on a scale
of a few Compton wavelengths, see e.g.\cite{IZ}. We may therefore define
the spinor wavefunctions 
\begin{eqnarray}
\Psi _{j,u,\theta }(x,t) &:&=\psi _{j}^{A}(x,t)\frac{\left| a_{j}^{\dagger
}(\theta )\right\rangle }{\sqrt{2\pi ^{2}p_{j}^{0}}}=\frac{e^{-i\vec{p}%
_{j}\cdot \vec{x}}}{\sqrt{2\pi p_{j}^{0}}}u_{j}(\theta ) \\
\Psi _{j,v,\theta }(x,t)^{\dagger } &:&=\psi _{j}^{A}(x,t)^{\dagger }\frac{%
\left| a_{j}^{\dagger }(\theta )\right\rangle }{\sqrt{2\pi ^{2}p_{j}^{0}}}=%
\frac{e^{-i\vec{p}_{j}\cdot \vec{x}}}{\sqrt{2\pi p_{j}^{0}}}v_{j}(\theta
)^{\dagger }\,.
\end{eqnarray}
With the help of these functions we obtain then for the defect system 
\begin{eqnarray}
\mathbf{\Psi }_{i,u,\theta }^{A}(x,t) &:&=\psi _{i}^{A}(x,t)\frac{\left|
a_{i,-}^{\dagger }(\theta )\right\rangle }{\sqrt{2\pi ^{2}p_{i}^{0}}}=\Theta
(-x)\left[ \Psi _{i,u,\theta }(x,t)+\Psi _{i,u,-\theta }(x,t)R_{i}^{\ast
}(\theta )\right]  \nonumber \\
&&+\Theta (x)T_{_{i}}^{\ast }(\theta )\left[ \Psi _{i,u,\theta }(x,t)+\Psi
_{i,u,-\theta }(x,t)\tilde{R}_{_{i}}^{\ast }(-\theta )\right]  \label{fx}
\end{eqnarray}
and the same function with $u\rightarrow v$. Since this expression resembles
a free wave, it can not be normalized properly and we have to localize the
wave in form of a wave packet by multiplying with an additional function, $%
\tilde{g}(p,t)$ in (\ref{free}) and its counterpart $g(x,t)$ in (\ref{fx}),
typically a Gau\ss ian. Then for the function $\mathbf{\Phi }_{i,u,\theta
}^{A}(x,t)=g(x,t)\mathbf{\Psi }_{i,u,\theta }^{A}(x,t)$, we can achieve that 
$\left\| \mathbf{\Phi }\right\| =1$.

\subsection{Harmonic spectra}

We are now in the position to determine the emission spectrum for which we
need to compute the absolute value of the Fourier transform of the dipole
moment 
\begin{equation}
\mathcal{X}_{j,u,\theta }(\Omega )=\left| \int_{0}^{\tau }dt\,\left\langle 
\mathbf{\Phi }_{j,u,\theta }^{A}(x,t)^{\dagger }x\mathbf{\Phi }_{j,u,\theta
}^{A}(x,t)\right\rangle \exp i\Omega t\right| \,\,.  \label{33}
\end{equation}
We localize now the wave packet in a region much smaller than the classical
estimate for the maximal amplitude the electron will acquire when following
the laser field. We achieve this with a Gau\ss ian $g(x,t)=\exp
(-x^{2}/\Delta )$, where $\Delta \ll eE_{0}/\omega ^{2}$.

\subsection{An example: Impurity of energy operator type}

As mentioned this type of defect, i.e.~$\mathcal{D}(\bar{\psi},\psi )=g\bar{%
\psi}\psi (x)$ can be obtained in a limit from the defect discussed in
section 2.2.1. Coupling the vector potential minimally to it yields 
\begin{equation}
\mathcal{D}_{AD}(\bar{\psi},\psi ,A_{\mu })=g\bar{\psi}(1+e/m\gamma ^{\mu
}A_{\mu })\psi \,,
\end{equation}
by invoking the equation of motion. 

\subsection*{2.7.1 Transmission amplitudes}

We can now determine the reflection and
transmission amplitudes as outlined above 
\begin{eqnarray}
R_{i}(\theta ,g,A/e,y) &=&\tilde{R}_{i}(\theta ,g,-A/e,-y)=R_{\bar{\imath}%
}(\theta ,g,A/e,-y)=\tilde{R}_{\bar{\imath}}(\theta ,g,-A/e,y)=\quad 
\nonumber \\
&&\frac{[y\dot{A}-\cosh \theta ]e^{-2iy\sinh \theta }}{[1-y\dot{A}\cosh
\theta ]-i\frac{g}{4}[\frac{4}{g^{2}}+1+A^{2}-y^{2}\dot{A}^{2}]\sinh \theta }%
\,.  \label{r1}
\end{eqnarray}
We denoted the differentiation with respect to time by a dot. The
transmission amplitudes turn out to be 
\begin{eqnarray}
T_{i}(\theta ,g,A/e,y) &=&\tilde{T}_{i}(\theta ,g,-A/e,-y)=T_{\bar{\imath}%
}(\theta ,g,-A/e,y)=\tilde{T}_{\bar{\imath}}(\theta ,g,A/e,-y)=\quad 
\nonumber \\
&&\frac{i\left[ 1-y^{2}\dot{A}^{2}+(A-\frac{2i}{g})^{2}\right] \sinh \theta 
}{\frac{4}{g}[1-y\dot{A}\cosh \theta ]-i[\frac{4}{g^{2}}+1+A^{2}-y^{2}\dot{A}%
^{2}]\sinh \theta }\,.  \label{t1}
\end{eqnarray}
Locating the defect at $\ y=0$, the derivative of $A$ does not appear
anymore explicitly in (\ref{r1}) and (\ref{t1}), such that it is clear that
this defect is of type I and admits an expansion of the form (\ref{exp}).
With the explicit expressions (\ref{r1}) and (\ref{t1}) at hand, we can
determine all the coefficients $t_{2k}(\theta )$ in (\ref{exp})
analytically. For this purpose let us first bring the transmission amplitude
into the more symmetric form 
\begin{equation}
\left\vert T_{i}(\theta ,g,A/e)\right\vert ^{2}=\frac{\tilde{a}_{0}(\theta
,g)+a_{2}(\theta ,g)A^{2}+a_{4}(\theta ,g)A^{4}}{a_{0}(\theta
,g)+a_{2}(\theta ,g)A^{2}+a_{4}(\theta ,g)A^{4}},  \label{tt}
\end{equation}
with 
\begin{eqnarray}
a_{0}(\theta ,g) &=&16g^{2}+(4+g^{2})^{2}\sinh ^{2}\theta ,\quad \quad 
\tilde{a}_{0}(\theta ,g)=(g^{2}-4)^{2}\sinh ^{2}\theta ,\qquad \\
\,\,a_{2}(\theta ,g) &=&2g^{2}(4+g^{2})\sinh ^{2}\theta ,\quad \quad \qquad
\ a_{4}(\theta ,g)=g^{4}\sinh ^{2}\theta .\qquad
\end{eqnarray}
We can now expand $\left\vert T(\theta ,g,A)\right\vert ^{2}$ in powers of
the field $A(t)$ and identify the coefficients $t_{2k}(\theta ,g)$ in (\ref%
{exp}) thereafter. To achieve this we simply have to carry out the series
expansion of the denominator in (\ref{tt}). The latter admits the following
compact form 
\begin{equation}
\frac{1}{a_{0}(\theta ,g)+a_{2}(\theta ,g)A^{2}+a_{4}(\theta ,g)A^{4}}%
=\sum_{k=0}^{\infty }c_{2k}(\theta ,g)A^{2k},
\end{equation}
with $c_{0}(\theta ,g)=1/a_{0}(\theta ,g)$ and 
\begin{equation}
c_{2k}(\theta ,g)=-\frac{c_{2k-2}(\theta ,g)a_{2}(\theta ,g)+c_{2k-4}(\theta
,g)a_{4}(\theta ,g)}{a_{0}(\theta ,g)},
\end{equation}
for $k>0$. We understand here that all coefficients $c_{2k}$ with $k<0$ are
vanishing, such that from this formula all the coefficients $c_{2k}$ may be
computed recursively. Hence, by comparing with the series expansion (\ref%
{exp}), we find the following closed formula for the coefficients $%
t_{2k}(\theta ,g)$%
\begin{equation}
t_{2k}(\theta ,g)=[\tilde{a}_{0}(\theta ,g)-a_{0}(\theta ,g)]c_{2k}(\theta
,g)\quad k>0.
\end{equation}
The first coefficients then simply read 
\begin{eqnarray}
t_{0}(\theta ,g) &=&\frac{\tilde{a}_{0}(\theta ,g)}{a_{0}(\theta ,g)}%
=|T(\theta ,E_{0}=0)|^{2}, \\
t_{2}(\theta ,g) &=&\frac{a_{2}(\theta ,g)}{a_{0}(\theta ,g)}\left[
1-t_{0}(\theta ,g)\right] =\frac{8g^{4}(4+g^{2})\sinh ^{2}2\theta }{%
(16g^{2}+(4+g^{2})^{2}\sinh ^{2}\theta )^{2}}, \\
t_{4}(\theta ,g) &=&\left[ \frac{a_{4}(\theta ,g)}{a_{2}(\theta ,g)}-\frac{%
a_{2}(\theta ,g)}{a_{0}(\theta ,g)}\right] t_{2}(\theta ,g),
\end{eqnarray}
and so on. It is now clear how to obtain also the higher terms analytically,
but since they are rather cumbersome we do not report them here.

\subsection*{2.7.2 Transmission amplitudes}

Having computed the coefficients $t_{2k}$, we can evaluate the series (\ref%
{ts}) and (\ref{tss}) in principle to any desired order. For some concrete
values of the laser and defect parameters the results of our evaluations are
depicted in figure 3.

\begin{figure}[h]
\epsfig{file=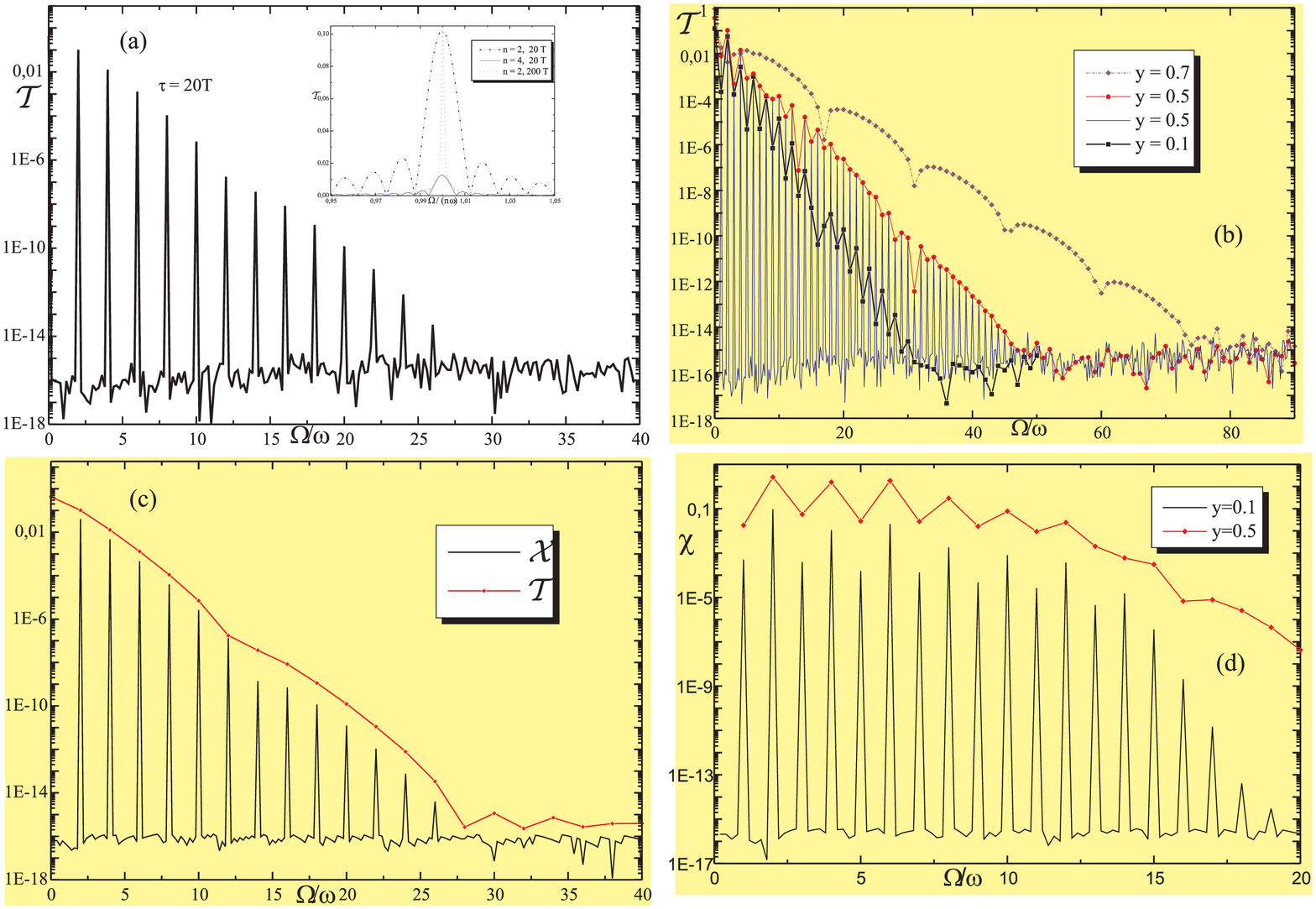,width=12.5cm,height=9.28cm} 
        \caption{Fourier transform of the transmission
probability for a single (a) and double (b) defect with $E_{0}=2.0$, $g=3.5$, 
$\theta =1.2$, $ \omega =0.2$.  Harmonic emission spectrum for a single (c) 
and double (d) defect with $E_{0}=2.0$, $g=3.5$, $\theta =1.2$, $\omega =0.2$, 
$\Delta =6$.  }    
\end{figure}

\noindent The main
 observation from part (a) is that the defect acts as a filter
selecting higher harmonics of even order of the laser frequency.
Furthermore, from the zoom of the peak regions, we see that there are
satellite peaks appearing near the main harmonics. They reduce their
intensity when $\tau $ is increased, such that with longer pulse length the
harmonics become more and more pronounced. We also investigated that for
different frequencies $\omega $ the general structure will not change.
Increasing the field amplitude $E_{0}$, simply lifts up the whole plot
without altering very much its overall structure. We support these findings
in two alternative ways, either by computing directly (\ref{TTT})
numerically or, more instructively, by evaluating the sums (\ref{ts}) and (%
\ref{tss}).

Part (b) shows the analysis for a double defect system with one defect
situated at $x=0$ and the other at $x=y$. The double defect amplitudes are
computed directly from (\ref{tr}) and (\ref{tr2}) with the expression for
the single defect (\ref{r1}) and (\ref{t1}). Since now both $A$ and $\dot{A}$
appear explicitly in the formulae for the $R$'s and $T$'s, it is clear that
the expansion of the double defect can not be of type I, but it turns out to
be of type II, i.e.~of the form (\ref{exp2}). Hence, we will now expect that
besides the even also the odd multiples of $\omega $ will be filtered out,
which is indeed visible in part (b) for various distances. Here we have only
plotted a continuous spectrum for $y=0.5$, whereas for reasons of clarity,
we only drew the enveloping function which connects the maxima of the
harmonics for the remaining distances. We observe that now not only odd
multiples of the frequency emerge in addition to the ones in (a) as
harmonics, but also that we obtain much higher harmonics and the cut-off is
shifted further to the ultraviolet. Furthermore, we observe a regular
pattern in the enveloping function, which appears to be independent of $y$.
Similar patterns were observed before in the literature, as for instance in
the context of atomic physics described by a Klein-Gordon formalism (see
figure 2 in\cite{Grob}).

Coming now to the main point of our analysis we would like to see how this
structure is reflected in the harmonic spectra. The result of the evaluation
of (\ref{33}) is depicted in figure 3 parts (c) and (d). We observe a very
similar spectrum as we have already computed for the Fourier transform of
the transmission amplitude, which is not entirely surprising with regard to
the expression (\ref{33}). The cut-off frequencies are essentially
identical. From the comparison between $\mathcal{X}$ and the enveloping
function for $\mathcal{T}$ we deduce, that the term involving the
transmission amplitude clearly dominates the spectrum.

The important general deduction from these computations is of course that 
\emph{harmonics of higher order do emerge in the emission spectrum of
impurity systems, such that harmonics can be generated from solid state
devices}.

\subsection*{2.7.3 Relativistic versus non-relativistic regime}

In the previous sections we have been working in an intensity regime which
is close to the damage threshold of a solid, according to the experimental
observations in\cite{solid1}. This allowed us to see the maximum effect
with regard to harmonic generation which at present might be visible from
experiments. However, it is also interesting to investigate situations which
are not experimentally feasible at present and of course lower intensity
regimes.

In order to judge in which regime we are working and whether there are
relativistic effects, let us carry out various limits. First of all we
recall a standard estimation according to which the relativistic kinetic
energy is close to the classical one when one is dealing with velocities 
$v^{2}\ll 3/4c^{2}$. This is the same as saying that the kinetic energy is
much smaller than the rest mass $E_{\rm kin }\ll m_{0}c^{2}$. Making now a
rough estimation for the system under consideration, we assume that the
total kinetic energy is the one obtained from the laser field, i.e. the
ponderomotive energy $U_{p}$. We also ignore for this estimation any
sophisticated corrections, such as possible Doppler shifts in the frequency,
etc. Then the non-relativistic regime is characterized by the condition $%
U_{p}\ll 1$.

Based on our previous observation that ${\cal T}$ and ${\cal X}$ exhibit a
very similar behaviour, it will be sufficient here to study only the ${\cal T%
}$ \ in the different regimes, which will be easier than an investigation of
the full emission spectrum (\ref{33}). From our analytic expression (\ref
{tss}), we see that for a type I defect the quantity ${\cal T}_{I}$ becomes
a function of $U_{p}$, such that the regime will be the same when we rescale
simultaneously $E_{0}$ and $\omega $. Accordingly we evaluate numerically
the Fourier transform (\ref{TT}), or equivalently compare against our
analytical expression (\ref{tss}), and depict our results in figure 4a.

\begin{figure}[h]
\epsfig{file=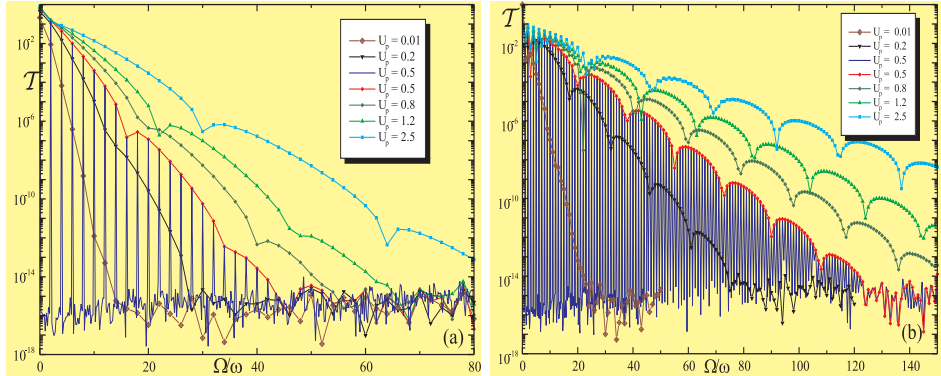,width=12.5cm,height=5.0cm} 
        \caption{Absolute value squared
of the Fourier transform of the transmission
probability. (a) Single defect for various values of $U_{p}$ 
with $g=3.5$, $\theta =1.2.$ (b) Double defect for varying values of 
$E_{0}$ with $g=3.5$, $\theta =1.2,\omega =0.2$.}    
\end{figure}

\noindent We observe 
that when passing more and more towards the relativistic regime
the cut-off is increased. The other feature one recognizes is that the
modulating structure in the enveloping function of the harmonics becomes
more pronounced. One should also note, in regard to (\ref{Dipole}), that the
multipole structures might become more and more important in the
relativistic regime.

Let us now perform a similar computation for the double defect. From the
expressions (\ref{II1}) and (\ref{II2}) we see that now ${\cal T}_{II}$ is
not just a simple function of $U_{p}$ and therefore even being in the same
regime the behaviour will be different when $E_{0}$ and $\omega $ are
rescaled. We alter in that case the regimes by rescaling $E_{0}$ and keeping
the frequency fixed. Our results are depicted in figure 4b.

Similar as for the single defect we see that the cut-off is increased and
the modulating structure in the enveloping function becomes more emphasized
when we move towards the relativistic regime. In addition we note that the
difference between the even and odd harmonic becomes larger with increasing $%
U_{p}$. This effect is more extreme for the low order harmonics.

As a general observation we state that there are not any effects which seem
to be special to the relativistic regime, but the transition to that regime
seems to be rather smooth.

We will now consider a quite different application, which allows, however,
to use integrability not only in a constraining but in a completely
constructive manner.

\section{Fractional quantum Hall systems}

The quantum\cite{Klitz} and in particular the fractional\cite{St} quantum
Hall effect have attracted an enormous amount of attention both from
theorist\cite{Laugh} and experimentalists (for some very recent experiments
see e.g.\cite{exp}). The key observation is that when subjecting an
electron gas confined to two space dimensions to a strong uniform magnetic
field, the transverse (Hall) conductance takes on preferably certain
characteristic values $G=e^{2}/h\nu $, whereas the longitudinal conductance
vanishes at these plateaux in complete analogy to the classical Hall 
effect\cite{Hall}. The filling fractions $\nu $ are distinct universal, in the
sense that they are independent of the geometry or type of the material,
rational numbers, which can be determined experimentally to an extremely
high precision. Many, but not all, of the experimentally observed filling
fractions are part of Jain's famous sequence (see\cite{Jain} and references
therein) 
\begin{equation}
\nu =\frac{m}{mp\pm 1}\qquad \quad m,p/2=1,2,3,\ldots  \label{Jain}
\end{equation}
which results as a theoretical prediction from a composite Fermion theory.

In the following we will show that these universal numbers also quantize the
conductance of quantum wires when described by minimal affine Toda field
theories\cite{ATFT}.

\subsection{Conductance in the high temperature regime}

There exist two established theoretical descriptions to compute the
conductance. The first is based on a linear response theory in which one
essentially needs the Fourier transform of the current-current two-point
correlation function. This so-called Kubo formula\cite{kubo,KTH} has been
adopted to a situation with a boundary\cite{LSS}. However, since this
set-up only captures effects coming from the constriction of the wire a
generalization which includes defects was needed, which we proposed in\cite%
{OA45} as

\begin{equation}
G^{\mathbf{\alpha }}(1/T)
=-\lim_{\omega \rightarrow 0}\frac{1}{2\omega \pi ^{2}%
}\int\nolimits_{-\infty }^{\infty }dt\,\,e^{i\omega t}\,\,\left\langle
J(t)Z_{\mathbf{\alpha }}\,J(0)\right\rangle _{T,m}.  \label{2}
\end{equation}
Here the defect operator $Z_{\mathbf{\alpha }}$ enters in-between the two
local currents $J$ within the temperature $T$ and mass $m$ dependent
correlation function. The Matsubara frequency is denoted by $\omega $.

The other possibility of determining the conductance is a generalization of
the Landauer-B\"{u}ttinger transport theory picture. Within this framework a
proposal for the conductance through a quantum wire with a defect (impurity)
has been made in\cite{FLS,FLS2} 
\begin{equation}
G^{\mathbf{\alpha }}(1/T)=\sum_{i}\lim_{(\mu _{i}^{l}-\mu _{i}^{r})\rightarrow
0}\frac{q_{i}}{2}\int\nolimits_{-\infty }^{\infty }d\theta \left[ \rho
_{i}^{r}(\theta ,T,\mu _{i}^{l})|T_{i}^{\mathbf{\alpha }}\left( \theta
\right) |^{2}-\rho _{i}^{r}(\theta ,T,\mu _{i}^{r})|\tilde{T}_{i}^{\mathbf{%
\alpha }}\left( \theta \right) |^{2}\right] ,  \label{11}
\end{equation}%
which we only modify to accommodate parity breaking\cite{OA45}. This means
we allow the transmission amplitudes for a particle of type $i$ with charge $%
q_{i}$ passing with rapidity $\theta $ through a defect of type $\mathbf{%
\alpha }$ from the left $T_{i}^{\mathbf{\alpha }}\left( \theta \right) $ and
right $\tilde{T}_{i}^{\mathbf{\alpha }}\left( \theta \right) $ to be
different. The density distribution function $\rho _{i}^{r}(\theta ,T,\mu
_{i})$ depends on the temperature $T$, and the potential at the left $\mu
_{i}^{l}$ and right $\mu _{i}^{r}$ constriction of the wire.

In both descriptions (\ref{2}) and (\ref{11}) one can employ
non-perturbative methods of integrable models, leading in (\ref{2}) to an
exact expression for the current-current correlation functions $\langle
\ldots \rangle _{T,m}$ from a form factor\cite{KW,Smir,BCFK} expansion and
in (\ref{11}) for the density distributions $\rho _{i}$ from a thermodynamic
Bethe ansatz\cite{TBAZam} analysis.

Here we will not consider (\ref{2}) and 
only present a more detailed study of (\ref{11}) when
impurities are absent and parity is preserved. In order to compute $G$ in a
one dimensional quantum wire, we simply have to determine the difference of
the static charge distribution at the left and right constriction of the
wire, which we assume to be at the potentials $\mu _{i}^{l}$ and $\mu
_{i}^{r}$, respectively. Then, to obtain the direct current $I_{i}$ for each
particle of type $i$ with charge $q_{i}$, we have to integrate the density
distribution functions $\rho _{i}^{r}(\theta ,T,\mu _{i})$ of occupied
states over the full range of the rapidities $\theta $ and the conductance
simply reads 
\begin{eqnarray}
G(1/T) &=&\sum_{i}\lim_{\Delta \mu _{i}\rightarrow 0}\frac{1}{\Delta \mu _{i}%
}I_{i}(1/T,\Delta \mu _{i}=\mu _{i}^{l}-\mu _{i}^{r}) \\
&=&\sum_{i}\lim_{\Delta \mu _{i}\rightarrow 0}\frac{q_{i}}{2\Delta \mu _{i}}%
\int\limits_{-\infty }^{\infty }d\theta \left[ \rho _{i}^{r}(\theta ,T,\mu
_{i}^{l})-\rho _{i}^{r}(\theta ,T,\mu _{i}^{l})\right] .  \label{1}
\end{eqnarray}%
Hence, the main task in this approach is to determine the density
distribution functions $\rho _{i}^{r}(\theta ,T,\mu _{i})$ of occupied
states. It is remarkable that in the context of integrable models, despite
the fact that these functions are neither Fermi-Dirac nor Bose-Einstein,
there exist approaches in which they can be computed non-perturbatively,
i.e.~the thermodynamic Bethe ansatz\cite{TBAZam}.

We briefly recall how this works. The central equations of the TBA
relate the total density of available states $\rho _{i}(\theta ,r)$ for
particles of type $i$ with mass $m_{i}$ as a function of the inverse
temperature $r=1/T$ to the density of occupied states $\rho _{i}^{r}(\theta
,r)$ 
\begin{equation}
\rho _{i}(\theta ,r)=\frac{m_{i}}{2\pi }\cosh \theta
+\sum\limits_{j}[\varphi _{ij}\ast \rho _{j}^{r}](\theta )\,.  \label{rho}
\end{equation}

\noindent By $\left( f\ast g\right) (\theta )$ $:=1/(2\pi )\int d\theta
^{\prime }f(\theta -\theta ^{\prime })g(\theta ^{\prime })$ we denote as
usual the convolution of two functions. There are only two inputs into the
entire TBA analysis: first\ the dynamical interaction, which enters via the
logarithmic derivative of the scattering matrix $\varphi _{ij}(\theta
)=-id\ln S_{ij}(\theta )/d\theta $ and an assumption on the statistical
interaction $g_{ij}$ amongst the particles $i$ and $j$ on which we comment
further below. For the moment we chose this interaction to be of fermionic
type. The mutual ratio of the two types of densities serves as the
definition of the so-called pseudo-energies $\varepsilon _{i}(\theta ,r)$%
\begin{equation}
\frac{\rho _{i}^{r}(\theta ,r)}{\rho _{i}(\theta ,r)}=\frac{e^{-\varepsilon
_{i}(\theta ,r)}}{1+e^{-\varepsilon _{i}(\theta ,r)}}\,,  \label{dens}
\end{equation}%
which have to be positive and real. At thermodynamic equilibrium they can be
computed from the non-linear integral equations 
\begin{equation}
rm_{i}\cosh \theta =\varepsilon _{i}(\theta ,r,\mu _{i})+r\mu
_{i}+\sum\limits_{j}[\varphi _{ij}\ast \ln (1+e^{-\varepsilon _{j}})](\theta
)\,,  \label{TBA}
\end{equation}%
where $r=m/T$, $m_{l}\rightarrow m_{l}/m$, $\mu _{i}\rightarrow \mu _{i}/m$,
with $m$ being the mass of the lightest particle in the model and chemical
potential $\mu _{i}<1$. As pointed out already in\cite{TBAZam} (here just
with the small modification of a chemical potential), the comparison between
(\ref{TBA}) and (\ref{rho}) leads to the useful relation 
\begin{equation}
\rho _{i}(\theta ,r,\mu _{i})=\frac{1}{2\pi }\left( \frac{d\varepsilon
_{i}(\theta ,r,\mu _{i})}{dr}+\mu _{i}\right) \,\sim \frac{1}{2\pi r}%
\epsilon (\theta )\frac{d\varepsilon _{i}(\theta ,r,\mu _{i})}{d\theta }.
\label{rhoe}
\end{equation}%
Here $\epsilon (\theta )=\Theta (\theta )-\Theta (-\theta )$ is the unit
step function, i.e.~$\epsilon (\theta )=1$ for $\theta >0$ and $\epsilon
(\theta )=-1$ for $\theta <0$. In equation (\ref{dens}), we assume that in
the large rapidity regime the density $\rho _{i}^{r}(\theta ,r,\mu _{i})$ is
dominated by the last expression in (\ref{rhoe}) and in the small rapidity
regime by the Fermi distribution function. Therefore, from (\ref{dens})
follows 
\begin{eqnarray}
\rho _{i}^{r}(\theta ,r,\mu _{i}) &=&\frac{e^{-\varepsilon _{i}(\theta
,r,\mu _{i})}}{1+e^{-\varepsilon _{i}(\theta ,r,\mu _{i})}}\rho _{i}(\theta
,r,\mu _{i}) \\
&\sim &\frac{1}{2\pi r}\epsilon (\theta )\frac{d}{d\theta }\ln \left[ 1+\exp
(-\varepsilon _{i}(\theta ,r,\mu _{i}))\right] \,\,.
\end{eqnarray}%
Using this expression in equation (\ref{1}), we can approximate the direct
current in the ultraviolet by 
\begin{equation}
\lim\limits_{r\rightarrow 0}I_{i}(r,\Delta \mu _{i})\sim \frac{q_{i}}{4\pi r}%
\int\limits_{-\infty }^{\infty }d\theta \ln \left[ \frac{1+\exp
(-\varepsilon _{i}(\theta ,r,\mu _{i}^{l}))}{1+\exp (-\varepsilon
_{i}(\theta ,r,\mu _{i}^{r}))}\right] \,\frac{d\epsilon (\theta )\,}{d\theta 
}\,,  \label{ya}
\end{equation}%
after a partial integration. Taking now the potentials at the end of the
wire to be $\mu _{i}^{r}=-\mu _{i}^{l}=\mu _{i}/2$ we carry out the limit $%
\Delta \mu _{i}\rightarrow 0$ in (\ref{1}) with the help l'Hospital's rule
and the conductance becomes 
\begin{equation}
\lim\limits_{r\rightarrow 0}G_{i}(r)\sim \frac{q_{i}}{2\pi r}%
\int\limits_{-\infty }^{\infty }d\theta \frac{1}{1+\exp [\varepsilon
_{i}(\theta ,r,0)]}\left. \frac{d\varepsilon _{i}(\theta ,r,\mu _{i}/2)}{%
d\mu _{i}}\right\vert _{\mu _{i}=0}\frac{d\epsilon (\theta )\,}{d\theta }\,.
\end{equation}%
Noting that $d\epsilon (\theta )/d\theta =2\delta (\theta )$, we obtain 
\begin{equation}
\lim\limits_{r\rightarrow 0}G_{i}(r)\sim \frac{q_{i}}{\pi r}\frac{1}{1+\exp
\varepsilon _{i}(0,r,0)}\left. \frac{d\varepsilon _{i}(0,r,\mu _{i}/2)}{d\mu
_{i}}\right\vert _{\mu _{i}=0}\,.  \label{g0}
\end{equation}%
The derivative $d\varepsilon _{i}(0,r,\mu _{i}/2)/d\mu _{i}$ can be obtained
by solving 
\begin{equation}
\frac{d\varepsilon _{i}(0,r,\mu _{i}/2)}{d\mu _{k}}=-\frac{r}{2}\delta
_{ik}+\sum_{j}N_{ij}\frac{1}{1+\exp \varepsilon _{j}(0,r,\mu _{i}/2)]}\frac{%
d\varepsilon _{j}(0,r,\mu _{j}/2)}{d\mu _{k}}\,,  \label{de}
\end{equation}%
which results from performing a constant TBA analysis on the $\mu _{k}$%
-derivative of (\ref{TBA}) in the spirit of\cite{TBAZam}. At this point
only the asymptotic phases of the scattering matrix enter via 
\begin{equation}
N_{ij}=\frac{1}{2\pi i}\lim_{\theta \rightarrow \infty }[\ln [S_{ij}(-\theta
)/S_{ij}(\theta )]]~.
\end{equation}%
In principle we have now all quantities needed to compute the conductance,
but to solve (\ref{de}) for the derivatives of the pseudo-energies is
somewhat cumbersome, see\cite{OA45,CF12} for such a computation.
Nonetheless, we can elaborate more on equation (\ref{de}) and simplify the
procedure further. For this purpose we introduce the quantity 
\begin{equation}
Y_{ij}:=\frac{1}{r(1+e^{\varepsilon _{i}})}\frac{d\varepsilon _{i}}{d\mu _{j}%
}~,
\end{equation}%
such that we can re-write equation (\ref{de}) equivalently as 
\begin{equation}
M_{ij}Y_{jk}=\frac{\delta _{ik}}{2}\quad \quad {\rm  with}\qquad 
M_{ij}:=N_{ij}-(1+e^{\varepsilon _{i}})\delta _{ij}  \label{M}
\end{equation}%
where the pseudoenergies satisfy the constant TBA equations 
\begin{equation}
e^{-\varepsilon _{i}}=\prod\limits_{j}(1+e^{-\varepsilon _{j}})^{N_{ij}}~.
\label{ctba}
\end{equation}%
Returning now to dimensional variables, i.e.~replacing $1/2\pi \rightarrow
e^{2}/h$, the conductance at high temperature in terms of the filling
fraction $\nu $ then simply results to 
\begin{equation}
G(0)=\frac{e^{2}}{h}\nu \quad \quad {\rm with} \qquad \nu =\frac{1}{2}%
\sum_{i,j}q_{i}(M^{-1})_{ij}~.  \label{nu}
\end{equation}%
This means we have reduced the entire problem to compute filling fractions
simply to the task of finding and inverting the matrix $M$. This is done in
two steps: First from the asymptotic phases of the scattering matrix we
compute $N_{ij}$ and subsequently we solve the constant TBA equations (\ref%
{ctba}). Then it is a simple matter of inverting the matrix (\ref{M}) and
performing the sums in (\ref{nu}).

In the context of the fractional quantum Hall effect one encounters very
often particles which obey some exotic (anyonic) statistics. So far we have
assumed our particles to obey fermionic type statistics as this choice is
most natural for the investigated theories\cite{TBAZam}. However, one can
easily implement more general statistics by adding a matrix $g_{ij}$ to the $%
N$-matrix\cite{BF}.

The formula (\ref{nu}) reminds of course on the well-known expressions for
the conductance as may be found for instance in\cite{FZ,Capp}. In that
context it was found\cite{FZ,WZ} that Jain's sequence (\ref{Jain}) can be
obtained simply from the $(m\times m)$-matrix 
\begin{equation}
M_{ij}=p\pm \delta _{ij}~.  \label{JJ}
\end{equation}
For this we have to take $q_{i}=2$ $\forall ~i$ in our expression (\ref{nu}%
). We will now demonstrate that the sequence (\ref{Jain}) can also be
obtained in a more surprising way from fairly complicated matrices, even
with non-rational entries, which result directly in the way indicated above,
namely from a TBA analysis of minimal affine Toda field theories\cite{ATFT}%
. Each Toda theory is associated to a Lie algebra \textbf{g} of rank $\ell $
and it is well known\cite{TBAKM} that in that case $N$ is an $(\ell \times
\ell )$-matrix which is of the general form 
\begin{equation}
N_{ij}=\delta _{ij}-2(K_{\mathbf{g}}^{-1})_{ij}~,  \label{NK}
\end{equation}
where $K_{\mathbf{g}}$ is the Cartan matrix related to \textbf{g} (see
e.g.~\cite{Hum}). The solutions to the constant TBA equations are also
known\cite{Resh,TBAKM} for most cases. In the ultraviolet limit these
theories possess Virasoro central charge $c=2\ell /(H+2)$, with $H$ being
the Coxeter number of the Lie algebra \textbf{g}.

\subsection{Fractional filling fractions from minimal affine Toda field
theory}

\noindent Here we only reproduce 
the most important and stable subsequence of Jain's
hierarchy, that is (\ref{Jain}) for $p=2$. 
Let us start with some concrete examples to
illustrate the working of our formulae. The first member of this series,
that is filling fraction $\nu =1/3$ is one of the best studied examples\cite%
{St,Laugh,nu13} of the fractional quantum Hall effect. We show now that this
particular filling fraction results in the ultraviolet limit from an $A_{3}$%
-affine Toda field theory. Specializing the general expression (\ref{NK}) to
the $A_{3}$-case, the solutions to the constant TBA equations (\ref{ctba})
are simply 
\begin{equation}
e^{\varepsilon _{1}}=e^{\varepsilon _{3}}=2,\qquad e^{\varepsilon _{2}}=3~.
\end{equation}
Then, the inverse of the $M$-matrix 
\begin{equation}
M_{ij}=\delta _{ij}-2(K_{A_{3}}^{-1})_{ij}-\delta _{ij}(1+e^{\varepsilon
_{i}})~
\end{equation}
is computed to 
\begin{equation}
M^{-1}=\frac{1}{36}\left( 
\begin{array}{ccc}
11 & -2 & -1 \\ 
-2 & 8 & -2 \\ 
-1 & -2 & 11%
\end{array}
\right) ~.
\end{equation}
From the fact that the $A_{\ell }$-minimal affine Toda field theories can
also be viewed as complex sine-Gordon models\cite{CSG}, we know\cite{DH}
that the charges in this theory are $q_{1}=q_{3}=1$, $q_{2}=2$, such that (%
\ref{nu}) yields 
\begin{equation}
\nu _{A_{3}}=1/3~.
\end{equation}
This is not entirely surprising as it is known\cite{Capp} that the
fractional quantum Hall effect with this filling fraction can be described
successfully in terms of a $c=1$ conformal field theory (CFT), as in the
case at hand. The next example, i.e.~$A_{5}$-minimal affine Toda field
theory, yields a less expected answer, even more since the $M$-matrix
contains non-rational entries. With (\ref{NK}) for $A_{5}$ the solutions to
the constant TBA equations are\cite{Resh,TBAKM} 
\begin{equation}
e^{\varepsilon _{1}}=e^{\varepsilon _{5}}=1+\sqrt{2},\qquad e^{\varepsilon
_{2}}=e^{\varepsilon _{4}}=2+2\sqrt{2},\quad \quad e^{\varepsilon _{3}}=3+2%
\sqrt{2}~.
\end{equation}
Assembling this into the $M$-matrix, it is clear that it will contain
non-rational entries. Evidently this matrix is not of the \ form (\ref{JJ})
and certainly falls out of the classification scheme based on integral
lattices\cite{FT}. Nonetheless, it will lead to a distinct rational value
for $\nu $. We compute the inverse of $M$ to 
\begin{equation}
M^{-1}=\left( 
\begin{array}{ccccc}
\left( \frac{35}{4}-6\sqrt{2}\right) & \left( {\frac{31}{2\,\sqrt{2}}}%
-11\right) & \left( {\frac{7-5\,\sqrt{2}}{4}}\right) & \left( 6-{\frac{17}{%
2\,\sqrt{2}}}\right) & \left( 3\,\sqrt{2}-{\frac{17}{4}}\right) \\ 
\left( {\frac{31}{2\,\sqrt{2}}}-11\right) & \left( 15-{\frac{21}{\sqrt{2}}}%
\right) & \left( {\frac{7\,\sqrt{2}-10}{4}}\right) & \left( 6\,\sqrt{2}-{%
\frac{17}{2}}\right) & \left( 6-{\frac{17}{2\,\sqrt{2}}}\right) \\ 
\left( {\frac{7-5\,\sqrt{2}}{4}}\right) & \left( {\frac{7\,\sqrt{2}-10}{4}}%
\right) & \left( {\frac{9}{4}}-{\frac{3}{\sqrt{2}}}\right) & \left( {\frac{%
7\,\sqrt{2}-10}{4}}\right) & \left( {\frac{7-5\,\sqrt{2}}{4}}\right) \\ 
\left( 6-{\frac{17}{2\,\sqrt{2}}}\right) & \left( 6\,\sqrt{2}-{\frac{17}{2}}%
\right) & \left( {\frac{7\,\sqrt{2}-10}{4}}\right) & \left( 15-{\frac{21}{%
\sqrt{2}}}\right) & \left( {\frac{31}{2\,\sqrt{2}}}-11\right) \\ 
\left( 3\,\sqrt{2}-{\frac{17}{4}}\right) & \left( 6-{\frac{17}{2\,\sqrt{2}}}%
\right) & \left( {\frac{7-5\,\sqrt{2}}{4}}\right) & \left( {\frac{31}{2\,%
\sqrt{2}}}-11\right) & \left( {\frac{35}{4}}-6\,\sqrt{2}\right)%
\end{array}
\right) ~.  \label{M5}
\end{equation}
Remarkably when taking into account that\cite{DH} $q_{1}=q_{5}=1$, $%
q_{2}=q_{4}=2$, $q_{3}=3$, we obtain by evaluating (\ref{nu}) for the matrix
(\ref{M5}) the simple ratio 
\begin{equation}
\nu _{A_{5}}=3/8~.
\end{equation}
We will now turn to the generic case. Taking the general solutions of the
constant TBA equations into account\cite{Resh,TBAKM} and using a generic
expression for the inverse of the Cartan matrix $K_{A_{\ell }}^{-1}=\min
(i,j)-ij/(\ell +1)$ in (\ref{NK}), the M-matrix for an $A_{2\ell +1}$%
-minimal affine Toda field theory can be written generically as 
\begin{equation}
M_{ij}=\frac{ij}{\ell +1}-2\min (i,j)-\delta _{ij}\frac{\sin \left( \frac{%
i\pi }{2\ell +4}\right) \sin \left( \frac{(i+2)\pi }{2\ell +4}\right) }{\sin
^{2}\left( \frac{\pi }{2\ell +4}\right) }~.  \label{Ml}
\end{equation}
As already indicated by the previous example this matrix is not of the form (%
\ref{JJ}) and does not fit into the classification scheme proposed in\cite%
{FT}. According to\cite{DH} we have the charges 
\begin{equation}
q_{i}=q_{2\ell +2-i}\qquad {\rm and} \qquad q_{i}=i\quad {\rm for }
\quad i\leq
\ell +1~.  \label{ql}
\end{equation}
As can be guessed from (\ref{M5}), it is not evident how to express the
inverse in terms of a simple closed expression. We can, however, invert (\ref%
{Ml}) case-by-case up to very high rank and we obtain from (\ref{nu})
together with (\ref{ql}) the sequence 
\begin{equation}
\nu _{A_{2\ell +1}}=\frac{\ell +1}{2\ell +4}~.  \label{JP22}
\end{equation}
Taking now $\ell =2m-1$, we obtain as a subsequence of this the most stable
part of Jain's sequence (\ref{Jain}) with $p=2$%
\begin{equation}
\nu _{A_{4m-1}}=\frac{m}{2m+1}~.  \label{Jp2}
\end{equation}
In summary: \emph{The conductance of a quantum wire which is described by a
massive }$A_{2\ell +1}$\emph{-minimal affine Toda field theory possesses in
the high temperature regime, in which the model turns into a conformal field
theory with Virasoro central }$c=(2\ell +1)/(\ell +2)$\emph{, a filling
fraction equal to (\ref{JP22}). In particular for }$\ell =2m-1$\emph{, we
obtain the Jain sequence (\ref{Jp2}).}

In order to illustrate the main idea we just presented here the most stable
of the Jain sequence. To see how to obtain other sequences we refer the
reader to\cite{CFHall}.

\section{Conclusions}

We have presented two concrete applications in which  non-perturbative
methods developed in the context of integrable quantum field theories can be
used to evaluate physical quantities. We predict the generation of harmonic
spectra from solid state devices and show that filling fractions occurring
in fractional quantum Hall systems can be obtained from minimal affine Toda
field theories.

\section*{Acknowledgements}

A.F. would like to thank the organizers for their kind invitation. We are
grateful to the Deutsche Forschungsgemeinschaft (Sfb288) for financial
support. We also acknowledge useful discussions with M.~Mintchev, E.~Ragoucy 
and P.~Sorba.

\end{document}